\documentclass[]{interact}

\usepackage{epstopdf}% To incorporate .eps illustrations using PDFLaTeX, etc.
\usepackage[caption=false]{subfig}% Support for small, `sub' figures and tables

\usepackage[numbers,sort&compress,merge]{natbib}% Citation support using natbib.sty
\bibpunct[, ]{[}{]}{,}{n}{,}{,}% Citation support using natbib.sty
\usepackage{graphicx}
\usepackage{multirow}
\usepackage{color}
\usepackage{amsmath}
\usepackage{multirow}
\usepackage{rotating}
\usepackage{pdflscape}
\usepackage{soul}
\usepackage{float}

\usepackage[utf8]{inputenc}

\theoremstyle{plain}% Theorem-like structures provided by amsthm.sty

\theoremstyle{definition}

\theoremstyle{remark}

\newcommand{\warn}[1]{#1}           %warning redfont
\newcommand{\supp}{\emph{Supplemental Material}}

\newcommand*{\plimsoll}{{\ensuremath{-\kern-4pt{\ominus}\kern-4pt-}}}

\def\parcder#1#2{\frac{\partial{#1}}{\partial{#2}}}

\def\parcder2#1#2{\frac{\partial^2{#1}}{\partial{#2}^2}}

\begin{document}

%\articletype{ARTICLE TEMPLATE}% Specify the article type or omit as appropriate

\title{Polyaromatic Hydrocarbons with an Imperfect Aromatic System as Catalysts of Interstellar H$_{2}$ Formation}

\author{
\name{Dávid P. Jelenfi$^{a,c}$, Anita Schneiker$^{a,b}$, Attila Tajti$^{c}$, Gábor Magyarfalvi$^{b}$, and György Tarczay$^{b,d}$}
\affil{$^{a}$ELTE Eötvös Loránd University, Hevesy György PhD School of Chemistry P. O. Box 32, H-1518, Budapest 112, Hungary;
$^{b}$ELTE – Eötvös Loránd University, Institute of Chemistry, Laboratory of Molecular Spectroscopy, PO Box 32, Budapest H–1518, Hungary;
$^{c}$ELTE - Eötvös Loránd University, Institute of Chemistry, Laboratory of Theoretical Chemistry, P. O. Box 32, H-1518, Budapest 112, Hungary;
$^{d}$MTA-ELTE Lendület Laboratory Astrochemistry Research Group, P. O. Box 32, H-1518, Budapest 112, Hungary}}

\maketitle

\begin{abstract}
Although H$_{2}$ is the simplest and the most abundant molecule in the Universe, its formation in the interstellar medium, especially in the photodissociation regions is far from being fully understood. According to suggestions, the formation of H$_{2}$ is catalyzed by polyaromatic hydrocarbons (PAHs) on the surface of interstellar grains. 
In the present study we have investigated the catalytic effect of small PAHs with an imperfect aromatic system. Quantum chemical computations were performed for the H-atom-abstraction and
H-atom-addition reactions of benzene, cyclopentadiene, cycloheptatriene, indene, and 1H-phenalene. Heights of reaction barriers and \warn{tunneling reaction rate constants} were computed with density
functional theory using the MPWB1K functional. For each molecule, the reaction path and the \warn{rate constants} were determined at 50 K using ring-polymer instanton theory, and the temperature dependence of the \warn{rate constants} was investigated for cyclopentadiene and cycloheptatriene. The computational results reveal that defects in the aromatic system compared to benzene can increase the rate of the catalytic H$_{2}$ formation at 50 K.
\end{abstract}


%%-----------------------------------------------
\section{Introduction}
\label{intro}

Since H$_{2}$ is the simplest and the most abundant molecule of the interstellar medium (ISM), the understanding its formation in detail is in the center of interest. 
A lot of questions related to this issue have arisen over the years, many of which are still not fully answered. 
Except for the very dense regions (e.g. stellar atmospheres, planetary nebulas), H$_{2}$ cannot form in the gas-phase of the ISM by three-body collisions due to the low particle density. 
The radiative association mechanism is not effective either due to the fact that the H$_{2}$ molecule has no dipole moment, and thus the reaction energy cannot be dissipated resulting in the H atoms colliding  without reaction  \cite{williams2013,Vidali2013,wakelam2017,rauls2008,bonfanti2011,barrales2018,cazaux2004,goumans2010,miksch2021,goumans2011,mennella2011,schneiker2020,haupa2019,schneiker2021,habart2003,habart2004,le2009,skov2014,petrie1992,scott1997,le1997,snow1998,barrales2019}.

It was first suggested by Salpeter and co-workers \cite{Gould1963,Hollenbach1971}, and it is now well accepted in astrochemistry that H$_{2}$ can form on the surface or in the ice layer of interstellar grains.
There are two basic surface reaction mechanisms: the Langmuir-Hinshelwood mechanism and the Eley-Rideal mechanism \cite{Vidali2013,wakelam2017}.
In the case of the former one, both H atoms are bound to the surface and at least one of them must be mobile. 
The problem with this mechanism is that it can only be effective in a certain temperature range (e.g. in the diffuse interstellar medium) because at very low temperatures the diffusion rate of H atoms in the ice is too low, whereas at higher temperatures the H atoms cannot stay on the surface long enough for an efficient H$_{2}$ formation. 
In the case of the Eley-Rideal mechanism one H atom is bound on the surface and then a gaseous H atom collides and reacts with this atom.
This mechanism requires a relatively high concentration of H atoms. 
Therefore, considering simple physisorption, it cannot be considered as a main interstellar H$_{2}$ formation mechanism either, especially above 20~K where the residence time of the H atoms is too short.

There are several complex solid-phase reaction mechanisms that might explain the formation rate of interstellar H$_{2}$ \cite{williams2013,Vidali2013,wakelam2017}.
The residence time of H atoms near the surface can be increased by capturing them in the pores of amorphous carbon or silicate grains. 
Another possibility is that H atoms are bound on carbon grains at a low temperature, and a sudden increase of the temperature (for example due to a supernova explosion) can lead to explosive recombination of H atoms in a runaway event. 
A third mechanism involves the chemisorption of H atoms on the surface of interstellar grains.

In the case of the latter mechanism, a molecule can first chemisorb an H atom which in the second step is abstracted by another H atom, resulting in the formation of an H$_{2}$ molecule and the reformation of the original catalyst molecule. 
Therefore, this process can be considered as a catalytic cycle. Being a cycle, the mechanism can also start with abstraction of a H atom, followed by an addition, resulting in the original molecule. 
Starting from a closed-shell molecule regardless to the first step being abstraction or addition, former studies revealed that this first step of the cycle usually has an activation energy  \cite{rauls2008,bonfanti2011,barrales2018,cazaux2004}. The second step is a reaction that includes a free radical and a H atom, and it is generally a barrierless process. 
Consequently, the rate-limiting process is the first step, which can take place via H-atom tunneling at low temperatures.

It was suggested on the basis of astronomical observations by Habart and co-workers that the high number of polycyclic aromatic hydrocarbons (PAHs) present in the photon-dominated regions (PDRs) may be related to the rapid formation of molecular hydrogen in these regions. 
Therefore, PAHs might act as catalysts in the H$_{2}$ formation in PDRs \cite{habart2003,habart2004,le2009,skov2014}. 
The potential catalytic role of various PAHs has already been investigated both experimentally and theoretically \cite{goumans2010,miksch2021,goumans2011,mennella2011,schneiker2020}.

Based on computations, the H-atom-addition reaction of benzene \cite{goumans2010,miksch2021} and pyrene \cite{goumans2011} have a reaction barrier that is permeable for H atoms by tunneling. Moreover, the reaction rates of these reactions are non-negligible at low temperatures in contrast to those of the H-atom-addition reactions of graphene or graphite. 
In addition, experiments have also proved that H atoms and H$_{2}$ molecules can react with benzene and small PAHs, forming so-called superhydrogenated species  \cite{petrie1992,scott1997,le1997,snow1998}. 
Another experimental evidence was provided by Menella and coworkers \cite{mennella2011}, showing that coronene can react with deuterium atoms in a D-atom-addition reaction and the resulting species is able to react with another deuterium atom, resulting in the formation of HD or D$_{2}$ and the reformation of coronene \cite{mennella2011}.
Therefore, this experiment demonstrated that neutral PAH molecules can act as catalysts in the formation of interstellar H$_{2}$.

Computations and experiments have also revealed that the carbon atoms at different positions in PAHs have different reactivity, and there is a specific order of sequences of hydrogenation of PAHs  \cite{jensen2019,cazaux2019}. 
It was shown that among the hydrogenated PAH (HPAH) isomers, the most stable is the one that contains the maximum possible number of non-hydrogenated aromatic rings \cite{Pla2020}.
Many types of defects in aromaticity can also lower the barrier of the H atom addition. 
For instance, oxygen functionalized PAHs, e.g. 6,13-pentacenequinone, has enhanced reactivity compared to PAHs \cite{jaganathan2022}. 
Barrales-Martínez and Gutiérrez-Oliva have performed computations for H-atom-addition reactions of some N-and Si-doped coronenes. 
They found that the H-atom-addition reaction onto the N atom located in the external ring position and onto the Si atom, regardless of its position, has no activation energy. 
In addition, H-atom-addition onto carbon atoms next to these heteroatoms was found to have a lower activation energy than for the H-atom-addition reaction of coronene and the corresponding \warn{rate constants are also larger} \cite{barrales2019}.
Miksch and co-workers investigated theoretically the rate of the H-atom-addition to benzene and small heterocycles, pyridine, pyrrole, furan, thiophene, silabenzene, and phosphorene, in a wide temperature range, from 50 to 500 K. 
According to their results, the H-atom-addition reaction rate onto carbon atoms next to the heteroatoms of pyrrole or furan can be ca. 200\,times faster than that onto the carbon atoms of benzene at a temperature of 50\,K. 
This is an indication that small heterocycles can be better catalysts than PAHs in the formation of interstellar H$_{2}$ \cite{miksch2021}. 
Several recent experiments on pyrrole and furan support these theoretical findings \cite{amicangelo2020, schneiker2022}. 
Schneiker et al. also investigated the reaction of 1H\nobreakdash-phenalene with H atoms \cite{schneiker2020}, 
finding that the H-atom-addition reaction of the phenalenyl radical is barrierless, whereas the H-atom-abstraction reaction of 1H\nobreakdash-phenalene has a very low barrier permeable for H atoms by H atom tunneling even at very low temperatures \cite{schneiker2020}.

In the present work, we investigated the influence of the aromatic character on the catalytic effect of PAHs. 
For this, the heights of reaction barriers, the reaction energies, and the \warn{rate constants} were computed for small PAHs with different, imperfect aromatic systems: cyclopentadiene, cycloheptatriene, indene, and 1H\nobreakdash-phenalene. 
Among these, cyclopentadiene and indene were recently identified in the ISM \cite{cernicharo2021, burkhardt2021}, while the possible formation mechanism of 1H\nobreakdash-phenalene under conditions relevant to the ISM was experimentally explored \cite{zhao2020}.
The results obtained for these model systems are compared to the same computational data obtained for benzene (identified in the ISM in 2001 \cite{cernicharo2001}). 
We also discuss how these PAH motifs can contribute to the interstellar H$_{2}$ formation.

%%---------------------------------------------------------------------------------
\section{Computational details}
\label{comput}
In this study, the reaction channels of the H\nobreakdash-atom\nobreakdash-abstraction and H\nobreakdash-atom\nobreakdash-addition reactions are examined
with density functional theory (DFT) and instanton ring polymer rate theory for small PAHs containing one saturated carbon atom. The results are compared with the corresponding reactions of benzene to investigate the role of the aromatic character in the catalytic effect on the hydrogen molecule formation. 
Goumans and co-workers \cite{goumans2010} investigated the hydrogenation reaction of benzene and found that computations with the MPWB1K functional \cite{mpwb1k} and double-$\zeta$ quality basis sets yield results comparable to high-level CCSD(T)/CBS computations.
Recently, Miksch and co-workers studied the hydrogenation reactions of small heterocycles and argue for a triple-$\zeta$ quality basis set which gives slightly lower \warn{rate constants} for benzene at low temperatures than the 6-31G*(*) basis originally used by Goumans et al.
Based on the findings of these studies \cite{goumans2010,miksch2021}, the MPWB1K functional was used with the cc-pVDZ and cc-pVTZ basis sets in our calculations. Since the MPWB1K/6-31G*(*) and the MPWB1K/def2-TZVP gave very similar results for benzene \cite{miksch2021}, the temperature dependence of the rate constant of the H\nobreakdash-atom\nobreakdash-addition reaction of benzene was not recomputed at MPWB1K/cc-pVTZ level.

For each molecule, the reaction channels were determined by transition state searches using the Gaussian 09 program package \cite{g09}, followed by the  validation of the path using intrinsic reaction coordinate (IRC) computations \cite{doi:10.1021/ar00072a001,Hratchian2005FindingMT}.

In the case of the monocyclic molecules cyclopentadiene and cycloheptatriene,  the H\nobreakdash-atom\nobreakdash-additions were investigated at all unsaturated carbon atoms (C2-C3 and C2-C4 positions, respectively) and the H\nobreakdash-atom\nobreakdash-abstraction at the saturated one (C1 position).
For the polycyclic indene and 1H-phenalene molecules we analysed H\nobreakdash-atom\nobreakdash-additions only at the unsaturated carbon atoms in the ring containing the saturated atom (C2 and C3 positions) and H\nobreakdash-atom\nobreakdash-abstraction only at the saturated carbon atoms (C1 position). 
The transition states and products were optimized at the MPWB1K/cc-pVTZ and MPWB1K/cc-pVDZ levels.
The barrier heights and reaction energies were computed as the energy difference between the transition state and the reactants and between products and reactants, respectively, with and without zero-point vibration energy (ZPVE) corrections.
Furthermore, the HOMA index \cite{HOMA} (Harmonic Oscillator Model of Aromaticity) was calculated for each species to characterize the aromaticity of these systems. 
The parameters for the HOMA index were chosen as $R_{opt} = 1.385$ \AA{} and $\alpha = 288.85$ \AA$^{-2}$, based on MPWB1K/cc-pVDZ calculations for benzene and cyclohexane.

At very low temperatures, classical transition state theory gives a wrong estimation of the \warn{rate constants} because below a characteristic temperature called the crossover temperature $T_c$, the tunneling mechanism becomes more favorable.
An approximation to the crossover temperature is given by \cite{gillan1987quantum}
\begin{equation}\label{eq:Tc}
    \centering
    T_c = \frac{\hbar\omega_i}{2\pi k_B},
\end{equation}
where $\omega_i$ is the absolute value of the imaginary frequency of the transition state.
The vibrational mode corresponding to this frequency determines the tunneling path.
\warn{In instanton ring polymer rate theory \cite{yairthesis,andersson2009comparison}, the \warn{rate constant} is given by the expression\cite{richardson2009ring}
\begin{equation}\label{eq:k}
    \centering
    k_{inst}  = \frac{1}{\beta_P \hbar} \sqrt{\frac{B_N}{2\pi\beta_P\hbar^2}} \frac{Q_{inst}}{Q_{reac}} e^{-S/\hbar},
\end{equation}
where $Q_{reac}$ and $Q_{inst}$ refer to the quantum partition functions of the reactants and the instanton, respectively, while $S$ is the so-called instanton action determined using a discretized closed Feynman path (CFP) integral\cite{callan1977,miller1975}. $B_N$ is a normalization factor and $\beta_P = \beta/P=(k_B T P)^{-1}$,  with $P$ being the number of \emph{beads} used in the calculation.
The beads are different replicas of the system from which the instanton is constructed, usually obtained from the harmonic expansion of the transition structure along the tunneling path \cite{callan1977,miller1975}.}
For a detailed introduction to instanton rate theory, its applications and capabilities, the reader is advised to study the excellent reviews of Litman \cite{yairthesis} and Richardson \cite{richardson2016microcanonical,richardson2018ring,richardson2009ring}.

\begin{figure}[H]
    \centering
    \includegraphics[width=0.5\textwidth]{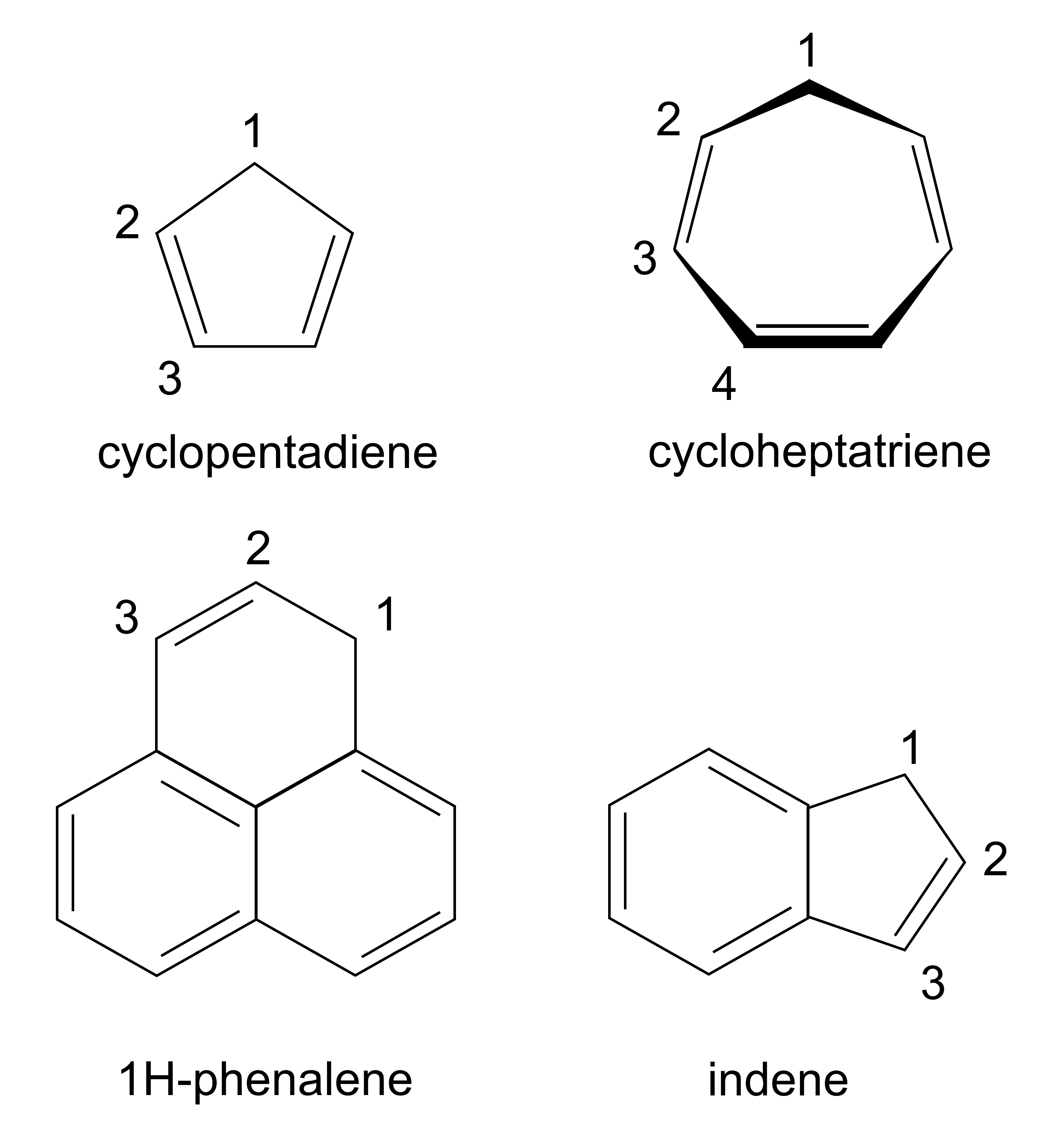}
    \caption{The investigated PAHs and the positions where the H-atom-abstraction and the H-atom-addition reactions were examined.}
    \label{fig:positions}
\end{figure}

The \warn{rate constants} were computed for the selected reaction channels using the i-PI software package \cite{i-PI} combined with FHI-aims program \cite{FHI-aims}.
The calculation of the \warn{rate constants} and their temperature dependence followed the procedure proposed by Litman \cite{yairthesis}.
This strategy uses a cooling procedure to determine the temperature dependence of the selected reactions, starting at a temperature around 20 K below $T_c$ and gradually decreased until 50 K.
At each temperature $T$, the instanton geometry is constructed from a gradually increasing number of beads $P$, determined by the formula suggested by Litman \cite{yairthesis}: \warn{$P \approx 40 \cdot T_c/T$}. 
(\warn{The exact number of utilized beads for each calculation can be found in the \supp.})
\warn{The initial instanton geometry was obtained from the transition state using a harmonic expansion along the imaginary vibrational mode, with the corresponding Hessian calculated analytically using the TURBOMOLE software package \cite{TURBOMOLE}. 
The optimization of the instanton geometry was performed 
%with the i-PI software package combined FHI-aims by 
using a first-order saddle point optimizer in the ring polymer potential \cite{nichols1990walking,richardson2012ring}.
For cyclopentadiene and cycloheptatriene the forces and the total energy of the instanton were converged to at least $10^{-5}$ eV/\AA$\,$ and $10^{-8}$ eV, respectively, while for the indene calculations $10^{-4}$ eV/\AA $\,$ and $10^{-7}$ eV were chosen as the respective convergency thresholds.
During this procedure the Hessian was updated according to the Powell formula \cite{fletcher2013practical}. 
Finally, the rate constants were determined from the action of the optimized instanton and its quantum partition function obtained from the analytical Hessian.} 
The reference (zero) energy of the PESs \warn{was} chosen as the electronic total energy of the reactants without ZPVE.
For reactions of the monocyclic molecules, the instanton geometry was optimized using the cc-pVTZ basis set. 
For the polycyclic molecules, due to the high cost of these computations, the instanton geometries were determined only at 50 K using the cc-pVDZ basis set. \warn{The rate constants for 1H-phenalene were not calculated because of the computational difficulties related to the size of this molecule.}

%%-----------------------------------------------
\section{Results and Discussion}

We start the discussion of the computational results with the  potential energy surfaces (PESs) of the first H-atom-abstraction and H-atom-addition reactions.
Both reactions result in a radical formation.
Usually, both the recombination reaction between this radical and an H atom and the H atom abstraction by an H atom from this radical is a barrierless process; therefore, we do not discuss these processes. 
In the next subsection we discuss the tunneling \warn{rate constants} obtained by instanton computations for 50 K.
In the third subsection the temperature dependence of these reactions are analyzed.
Finally, in the last subsection, we discuss the correlation between \warn{rate constants}, barrier heights, and the HOMA indices.

\subsection{Potential energy surfaces}
\begin{figure}[H]
    \centering
    \includegraphics[width=14cm]{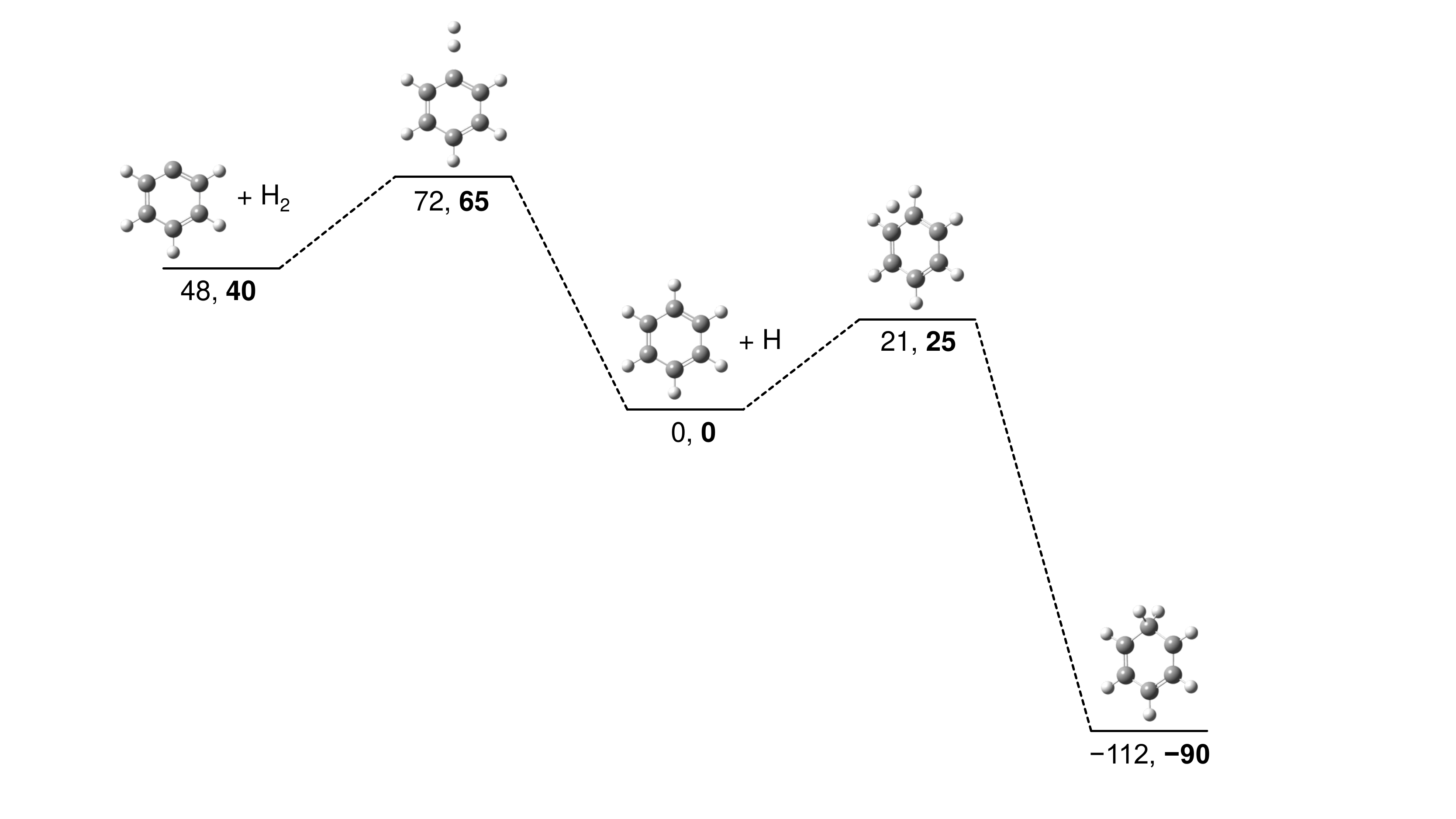}
    \caption{PES of the first H-atom-addition reaction and the first H-atom-abstraction reaction of benzene as computed at the MPWB1K/cc-pVTZ level of theory without (normal characters) and with (bold characters) ZPVE correction. Values are in kJ mol$^{-1}$.}
    \label{fig:benzene_barrier}
\end{figure}

\begin{figure}[H]
    \centering
    \includegraphics[width=14cm]{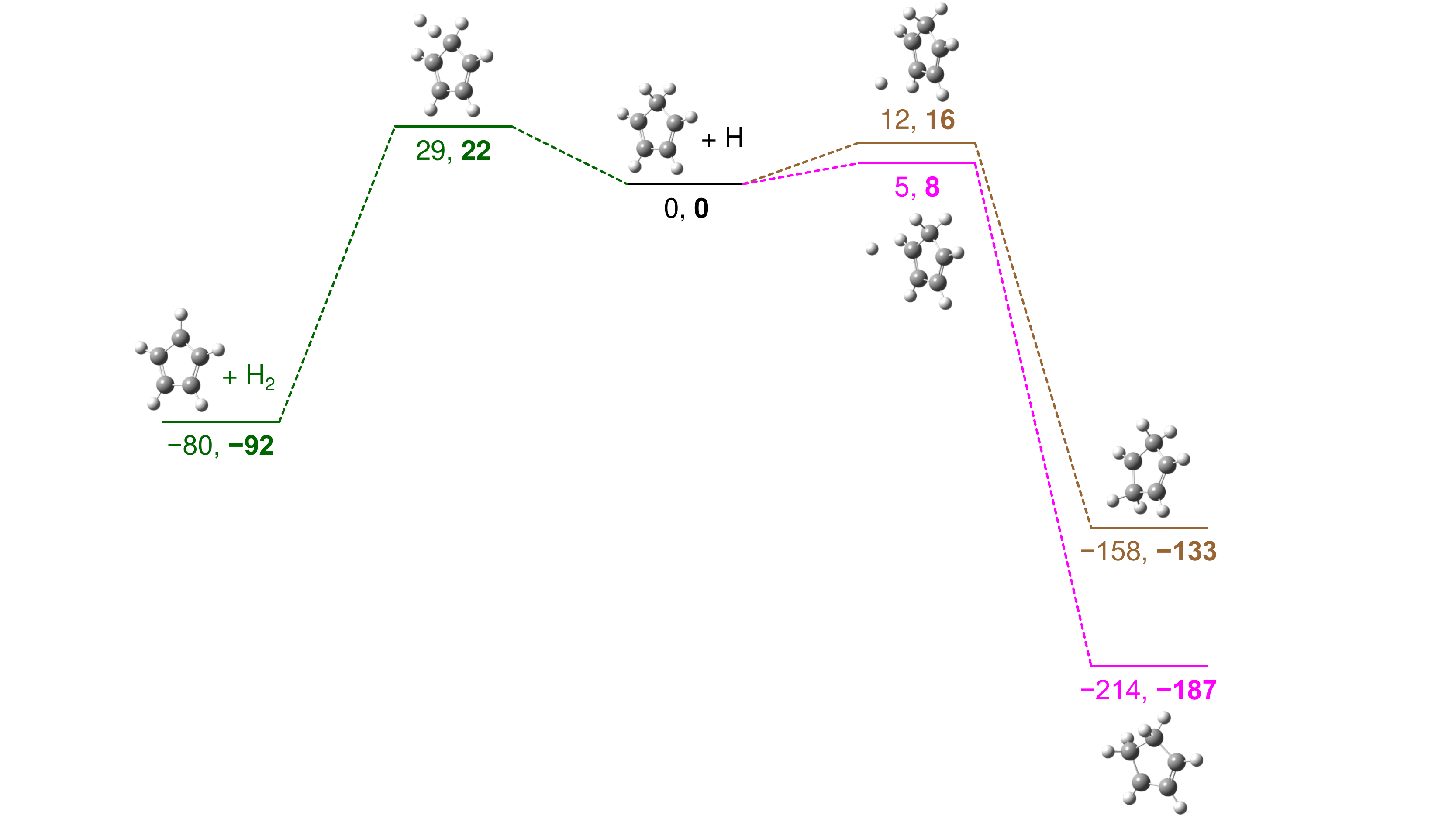}
    \caption{PES of the first H-atom-addition reactions and the first H-atom-abstraction reaction of cyclopentadiene as computed at the MPWB1K/cc-pVTZ level of theory without (normal characters) and with (bold characters) ZPVE correction. Values are in kJ mol$^{-1}$.}
    \label{fig:cp_barrier}
\end{figure}

\begin{figure}[H]
    \centering
    \includegraphics[width=14cm]{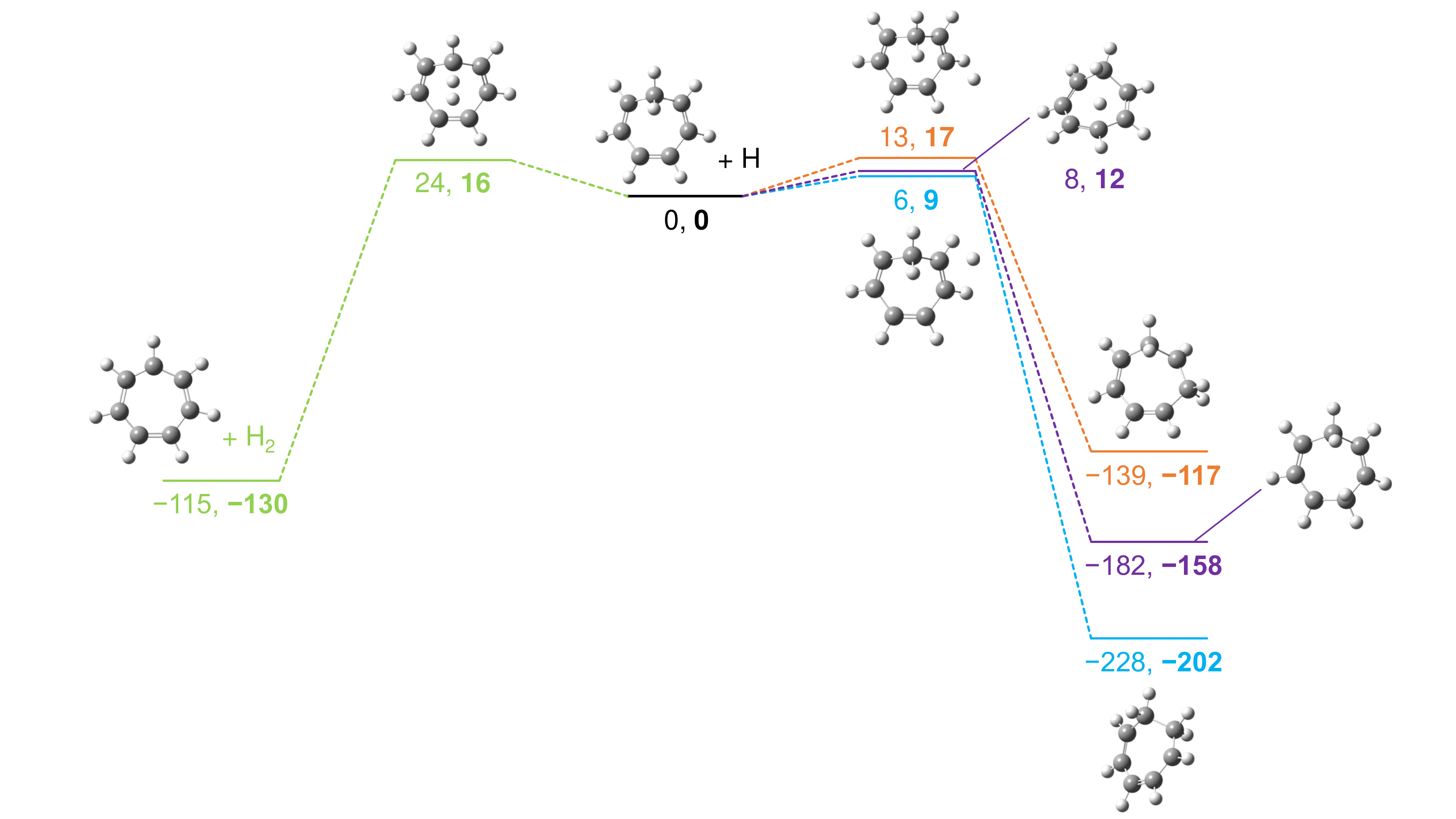}
    \caption{PES of the first H-atom-addition reactions and the first H-atom-abstraction reaction of cycloheptatriene via the upper path (see text) as computed at the MPWB1K/cc-pVTZ level of theory without (normal characters) and with (bold characters) ZPVE correction. Values are in kJ mol$^{-1}$.}
    \label{fig:cht_barrier_A}
\end{figure}

\begin{figure}[H]
    \centering
    \includegraphics[width=14cm]{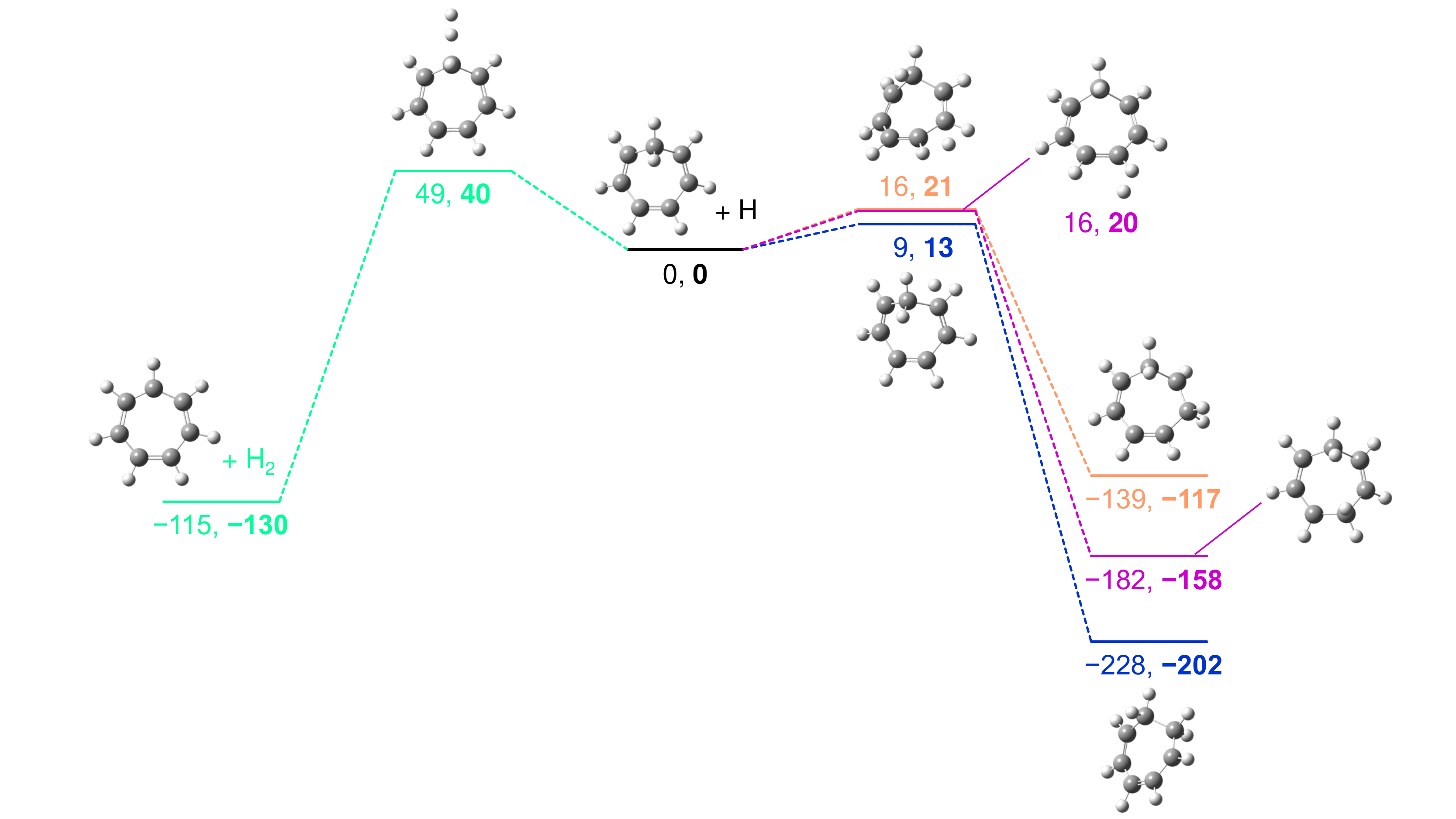}
    \caption{PES of the first H-atom-addition reactions and the first H-atom-abstraction reaction of cycloheptatriene via the lower path (see text) as computed at the MPWB1K/cc-pVTZ level of theory without (normal characters) and with (bold characters) ZPVE correction. Values are in kJ mol$^{-1}$.}
    \label{fig:cht_barrier_B}
\end{figure}

\begin{figure}[H]
    \centering
    \includegraphics[width=14cm]{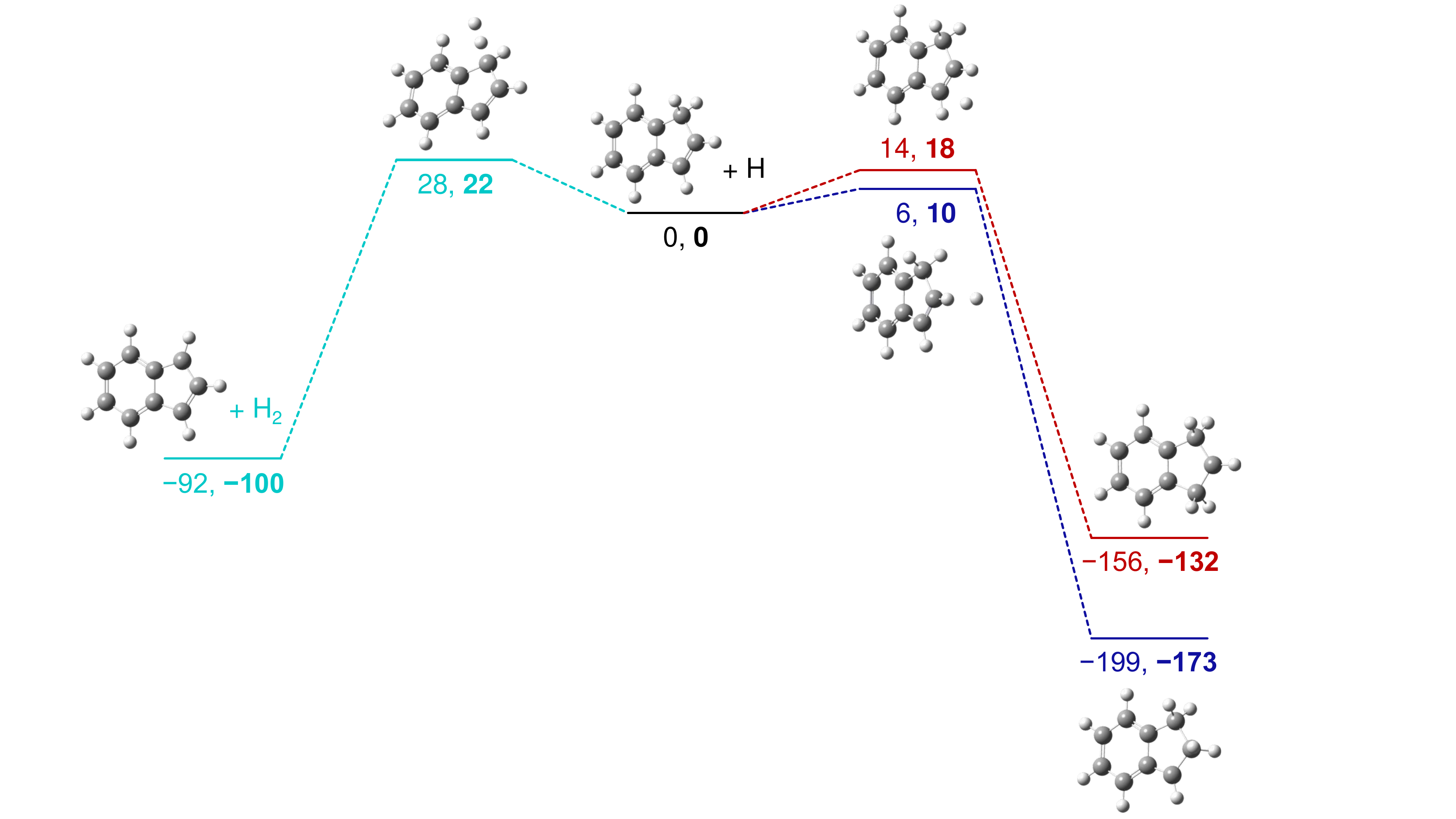}
    \caption{PES of the first H-atom-addition reactions and the first H-atom-abstraction reaction of indene as computed at the MPWB1K/cc-pVTZ level of theory without (normal characters) and with (bold characters) ZPVE correction. Values are in kJ mol$^{-1}$.}
    \label{fig:ind_barrier}
\end{figure}

\begin{figure}[H]
    \centering
    \includegraphics[width=14cm]{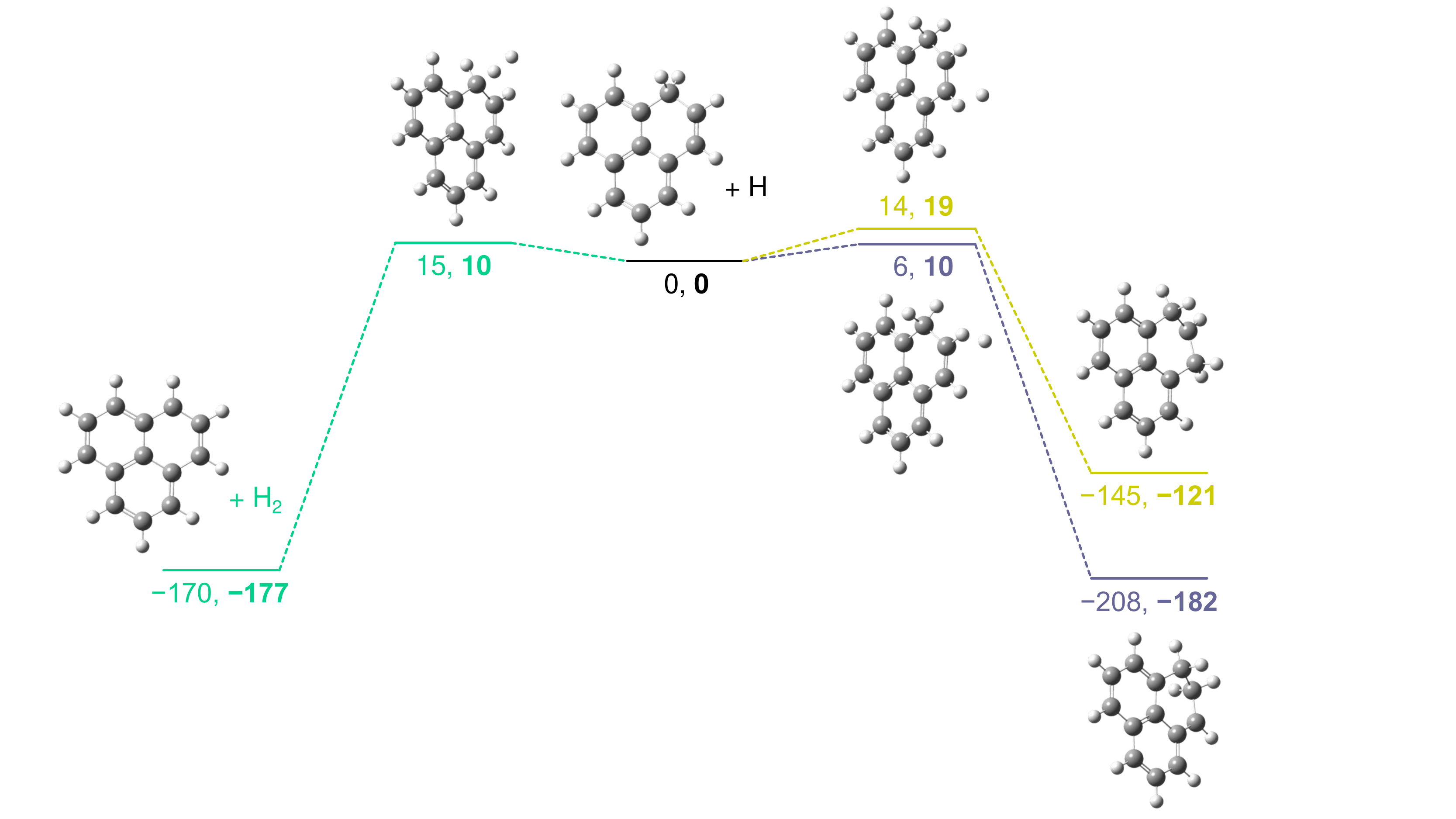}
    \caption{PES of the first H-atom-addition reactions and the first H-atom-abstraction reaction of 1H-phenalene as computed at the MPWB1K/cc-pVTZ level of theory without (normal characters) and with (bold characters) ZPVE correction. Values are in kJ mol$^{-1}$.}
    \label{fig:ph_barrier}
\end{figure}

The first H-atom-abstraction and H-atom-addition PESs of benzene, cyclopentadiene, cycloheptatriene, indene and 1H-phenalene, computed at the MPWB1K/cc-pVTZ level of theory, are illustrated in Figures \ref{fig:benzene_barrier},  \ref{fig:cp_barrier},  \ref{fig:cht_barrier_A}--\ref{fig:cht_barrier_B},  \ref{fig:ind_barrier}, and  \ref{fig:ph_barrier}, respectively.

Benzene, being a perfect aromatic system, was considered as the reference molecule. 
As it is shown in Figure \ref{fig:benzene_barrier}, the addition of the first H atom to benzene has a barrier of 21 kJ mol$^{-1}$ and 25 kJ mol$^{-1}$ at the MPWB1K/cc-pVTZ level of theory without and with ZPVE correction, respectively. 
These results are in agreement with the computational results of Goumans and Kästner \cite{goumans2010} and also with those of Miksch and co-workers \cite{miksch2021}.
In the former study the barrier height, computed at the MPWB1K/6-31G*(*) level including ZPVE was found to be 23.7 kJ mol$^{-1}$, while in the latter one 25.8 kJ mol$^{-1}$ was obtained at the MPWB1K/def2-TZVP level of theory. 
All these results are also in good agreement with the 22.9 kJ mol$^{-1}$ value obtained at the higher, CCSD(T)/CBS level of theory including the ZPVE correction.

Compared to the H-atom-addition reaction, the H-atom-abstraction reaction of benzene has a considerably larger barrier, 72 and 65 kJ mol$^{-1}$ at the MPWB1K/cc-pVTZ level of theory without and with ZPVE correction, respectively. 
In addition, the computed reaction enthalpy at 0 K of the H-atom-abstraction reaction of benzene was determined to be +48 and +40 kJ mol$^{-1}$ without and with ZPVE correction, respectively.
Since this reaction is endothermic, it cannot proceed at low temperatures, thus it has no astrochemical relevance.

The first H-atom-addition reactions of cyclopentadiene have smaller barriers than that of benzene (Figure \ref{fig:cp_barrier}). 
The H-atom-addition to C2 has a barrier of 5 and 8 kJ mol$^{-1}$ without and with ZPVE correction, respectively, while the H-atom-addition to C3 has a somewhat higher barrier,  12 and 16 kJ mol$^{-1}$ without and with ZPVE correction, respectively. 
This difference between the C2 and C3 positions can be explained by the larger electron delocalization in the radical formed by a H-atom-addition at C2. 
Both of these barriers are considerably smaller than that of benzene.

The H-atom-abstraction reaction of cyclopentadiene from the sp$^{3}$ C atom has a barrier of 29 and 22 kJ mol$^{-1}$ without and with ZPVE correction, respectively. 
These values are much smaller than the corresponding values for benzene and the ZPVE corrected value is slightly smaller even than the computed barrier height of the H-atom-addition reaction of benzene.
Moreover, the enthalpy of the reaction at 0 K for the H-atom-abstraction reaction of cyclopentadiene is as low as --80 and --92 kJ mol$^{-1}$ without and with ZPVE correction, respectively.
Consequently, in contrast to benzene, purely considering the energetics, both the H-atom-abstraction and the H-atom addition reactions are feasible for cyclopentadiene at low temperatures.

Cycloheptatriene has a planar and a boat form \cite{saebo1982}.
According to former theoretical studies \cite{saebo1982} and experiments \cite{traetteberg1964,butcher1965} the boat form has lower energy.
Therefore, in this work we investigated the reactions of the boat form only.
H atoms can react with the boat form via two different paths.
We refer to the \emph{upper path} when a H atom approaches the ring from the side where the sp$^{3}$ C atom sticks out from the plane.
Contrastingly, if the 
H atom approaches the ring from the opposite direction, we will refer to it as the \emph{lower path}.

The barrier heights of the H-atom-addition reaction of cycloheptatriene on C2 for the upper path are 6 and 9 kJ mol$^{-1}$ without and with ZPVE correction, respectively (Figure \ref{fig:cht_barrier_A}).
The barrier is somewhat higher for the lower path  (9 and 13 kJ mol$^{-1}$ without and with ZPVE correction, respectively, Figure \ref{fig:cht_barrier_B}).
The H-atom-addition reaction on C3 for the upper path (Figure \ref{fig:cht_barrier_A}) has a barrier of 13 and 17 kJ mol$^{-1}$ without and with ZPVE correction, respectively, while the corresponding values for the lower path (Figure \ref{fig:cht_barrier_B}) are 16 and 21  kJ mol$^{-1}$. 
Finally, when an H atom approaches the ring at C4, barriers of 8 and 12 kJ mol$^{-1}$ and 16 and 20 kJ mol$^{-1}$ are found for the upper (Figure \ref{fig:cht_barrier_A}) and lower paths (Figure \ref{fig:cht_barrier_B}), without and with ZPVE correction, respectively.
(Note that regardless of whether the H atom approaches the ring from the upper or from the lower direction, the products are identical, thus these reactions differ from each other only in the structure of the transition states.)
Considering the barrier height only, all these H-atom-addition reactions of cycloheptatriene are more favorable than that of benzene.
In general, the reactions in which the H atom approaches the ring from the upper side are more favorable than those with the H atom approaching from the lower direction.

The H-atom-abstraction reaction of cycloheptatriene from the sp$^{3}$ C atom has barriers of 24 and 16 kJ mol$^{-1}$ for the upper path (Figure \ref{fig:cht_barrier_A}) and 49 and 40 kJ mol$^{-1}$ for the lower path (Figure \ref{fig:cht_barrier_B}), without and with ZPVE correction, respectively.
Although, in contrast to benzene, these reactions are  exothermic, the lower-path reaction has too high of a barrier to be relevant at low temperatures.

Indene and 1H-phenalene were investigated from among polycyclic species. 
Indene contains a six- and a five-membered ring and the sp$^{3}$ C atom is part of the 5-membered ring. 
The barrier heights of the first H-atom-addition reactions of indene are both lower than that of benzene.
On C2, a barrier of 6 and 10 kJ mol$^{-1}$ is seen, without and with ZPVE correction, respectively, whereas 14 and 18 kJ mol$^{-1}$ are found for the corresponding values in the C3 position (Figure \ref{fig:ind_barrier}).
Similarly to cyclopentadiene, we can rationalize the energy difference of the two paths by the more extensive delocalization in the radical formed in the C2 addition. 
Nevertheless, both H-atom-addition reactions of indene have considerably lower barriers than that of benzene.

The barrier height of the H-atom-abstraction reaction of indene from the sp$^{3}$ C atom is 28 and 22 kJ mol$^{-1}$ without and with ZPVE correction, respectively (Figure \ref{fig:ind_barrier}). 
In contrast to benzene, this reaction is exothermic, the reaction enthalpy at 0 K being --92 and --100 kJ mol$^{-1}$ without and with ZPVE correction, respectively.
Therefore, the H-atom-abstraction reactions of indene might be relevant at low temperatures.

The largest molecule in our present investigation is 1H-phenalene, consisting of three six-membered rings.
1H-phenalene, sometimes referred simply as phenalene, is the lowest-energy one among the isomers of 1H-phenalene \cite{zoellner2008}.
The first H-atom-addition reaction of this system on C2 has a barrier of 6 and 10 kJ mol$^{-1}$ and 14 and 19 kJ mol$^{-1}$ on C3, without and with ZPVE correction, respectively (Figure \ref{fig:ph_barrier}).
Similarly to the other molecules discussed above, the difference in the barrier heights can again be explained by the more extensive electron delocalization in the radical formed in the reaction on C2. 
Both reactions possess, according to our computational results, a lower barrier than the corresponding process of benzene.

Regarding the first H-atom-abstraction from the sp$^{3}$ C atom, a barrier of 15 and 10 kJ mol$^{-1}$ is observed for 1H-phenalene without and with ZPVE correction, respectively. 
This reaction is highly exothermic, the reaction enthalpy being --170 and --177 kJ mol$^{-1}$ without and with ZPVE correction, respectively.
Among the investigated reactions in this study, this process has one of the the lowest barriers.
This can be explained by the fact that the phenalenyl radical, formed in the first H-atom-abstraction reaction of 1H-phenalene, is very stable due to the strong resonance stabilization \cite{gd2011}.
This radical can be prepared even in solution at room temperature \cite{goto1999}.

As a summary, we can conclude that according to our computational results, the molecules with imperfect aromatic systems have generally lower barriers for the H-atom-addition and the H-atom-abstraction reactions than benzene. 

\subsection{\warn{Rate constants} at 50 K}

At low temperatures, the investigated H-atom-addition and the H-atom-abstraction reactions can only proceed via tunneling.
Since the rate of tunneling reactions is determined rather by the width of the barrier than by its height, to get an insight into the catalytic activity in the low-temperature regime, one has to compute and analyze the \warn{rate constants}.
The \warn{rate constants} predicted by instanton theory at 50 K at the MPWB1K/cc-pVDZ level of theory, together with some computed literature values for benzene and other molecules, are listed in Table \ref{tab:50K_instanton}. 
As it can be seen, all considered reactions are faster by at least an order of magnitude than the H-atom-addition of benzene\warn{, except the H-atom-abstraction of the cyclopentadiene and the indene, as well as the H-atom-addition to C3 position of cycloheptatriene. In the latter cases, the rate constant only slightly differs from the rate constant of H-atom-addition of benzene.}
\warn{The fastest reactions, the H-atom-additions to C2 positions on cyclopentadiene, and on indene are about three orders of magnitude faster at 50 K than the H-atom-addition of benzene, and two orders of magnitude faster than the reaction of furan at C2 \cite{miksch2021}.}
These data, together with those of Miksch et al. \cite{miksch2021} show that the most favorable position for the H-atom-addition reaction is the carbon atom adjacent to the sp$^3$ carbon or, in the case of heterocycles, the one adjacent to the heteroatom.
This clearly indicates the role of aromaticity defects in the catalytic effect.

\begin{table}[H]
    \centering
    \caption{\warn{Reaction rate constants ($k$)} at 50 K temperature, barrier heights$^a$ $\Delta E^{\#}$, reaction enthalpies$^a$ at 0 K $\Delta_{r}H_{0\text{K}}^\plimsoll$, and the $\Delta$HOMA aromaticity indices$^b$ of H-atom-addition and H-atom-abstraction reactions for the PAHs investigated in the present study, computed at the MPWB1K/cc-pVDZ level of theory.
    Literature computational data are also listed for benzene, pyrene, and furan.}
    \vspace{1ex}
    {\small
    \begin{tabular}{|c|l|r|r|r|r|}
     \hline
       & \footnotesize{Reaction} & 
         \footnotesize{log ($k$ / cm$^3$s$^{-1}$)} &
         \footnotesize{$\Delta$HOMA$^b$} &
         \footnotesize{$\Delta E^{\#}$ / kJmol$^{-1}$} & \footnotesize{$\Delta_{r}H_{0\text{K}}^\plimsoll$ / kJmol$^{-1}$} \\
     \hline
      \multirow{4}{*}{benzene}
        & Abstraction from C1   & - & -0.03 & 73.4 & 55.8 \\ %TZ struct,TZ energies
        & Addition to C1 & -17.59 & -1.12 & 21.4 & -110.5 \\ 
        & Addition to C1$^c$ & -18.70 & - & 23.7 &  - \\ 
        & Addition to C1$^d$ & -19.00 & - & 25.8 & -89.2 \\ 
    
         \hline
       \multirow{3}{*}{pyrene}
        & Addition to C1$^e$      & -16.5 & - & 12.1 & - \\
        & Addition to C4$^e$      & -17.0 & - & 13.4 & - \\
        & Addition to C3a$^{1e}$     & -22.4 & - & 37.3 & - \\
        
        \hline
      \multirow{3}{*}{cyclopentadiene}
        & Abstraction from C1 & \warn{-18.07} & 1.11 & 27.5 & -72.7  \\ 
        & Addition to C2      & \warn{-13.93} & -0.90 & 4.7  & -213.2 \\ 
        & Addition to C3      & \warn{-16.32} & -1.02 & 12.3 & -157.7 \\ 
     \hline
       \multirow{4}{*}{cycloheptatriene}
        & Abstraction from C1$^f$ & \warn{-16.79} & 1.03 & 21.5 & -109.1 \\ %50K TZ k
        & Addition to C2$^f$      & \warn{-15.61} & -0.37 & 5.8 & -229.8 \\ %50K TZ k
        & Addition to C3$^f$      & \warn{-17.93} & -0.82 & 13.3 & -139.6 \\ %50K TZ k
        & Addition to C4$^f$      & \warn{-16.28} & -0.85 & 8.3 & -181.3 \\ %50K TZ k
     \hline
       \multirow{3}{*}{indene}
        & Abstraction from C1 & \warn{-17.50} &  0.66 & 27.8 & -84.6 \\ 
        & Addition to C2      & -15.12 & -0.50 & 6.4 & -198.5 \\ 
        & Addition to C3      & -16.96 & -0.44 & 14.1 & -155.8 \\ 
%     \hline
%       \multirow{3}{*}{1H-phenalene}
%        & Abstraction from C1 & -14.70 & 0.60 & 13.7 & -162.0 \\ %expected k
%        & Addition to C2      & -14.80 & -0.17 & 6.3 & -208.3 \\ %expected k
%        & Addition to C3      & -17.00 & -0.24 & 14.7 & -144.5 \\ %expected k
     \hline
       \multirow{2}{*}{furan}
        & Addition to C2$^d$      & -16.80 & - & 15.3 & -132.1 \\ 
        & Addition to C3$^d$      & -19.85 & - & 27.3 &  -83.2 \\

     \hline
    \end{tabular}}
    \label{tab:50K_instanton}
    
    \raggedright
    \footnotesize{$^a$ The barrier heights and reaction energies were computed as the energy difference between the transition state and the reactants and between products and reactants, respectively, without ZPVE corrections.}
    
    \footnotesize{$^b$ Change in the HOMA aromaticity index upon the reaction, calculated by subtracting the HOMA index of the reactant from the HOMA index of the product.}
    
    \footnotesize{$^c$ MPWB1K/6-31G*(*) data from ref. \citep{goumans2010}}.
    
    \footnotesize{$^d$ MPWB1K/def2-TZVP data from ref. \citep{miksch2021}}.
    
    \footnotesize{$^e$ MPWB1K/6-31G*(*) data from ref. \citep{goumans2011hydrogen}. (Approximate reaction rate values were read from Figure 2 within.)}
    
    \footnotesize{$^f$ Data for the \emph{upper path} (see text for details)}.

\end{table}

\subsection{Temperature dependence of the \warn{rate constants}}

\begin{figure}[H]
    \centering
    \includegraphics[width=0.5\textwidth]{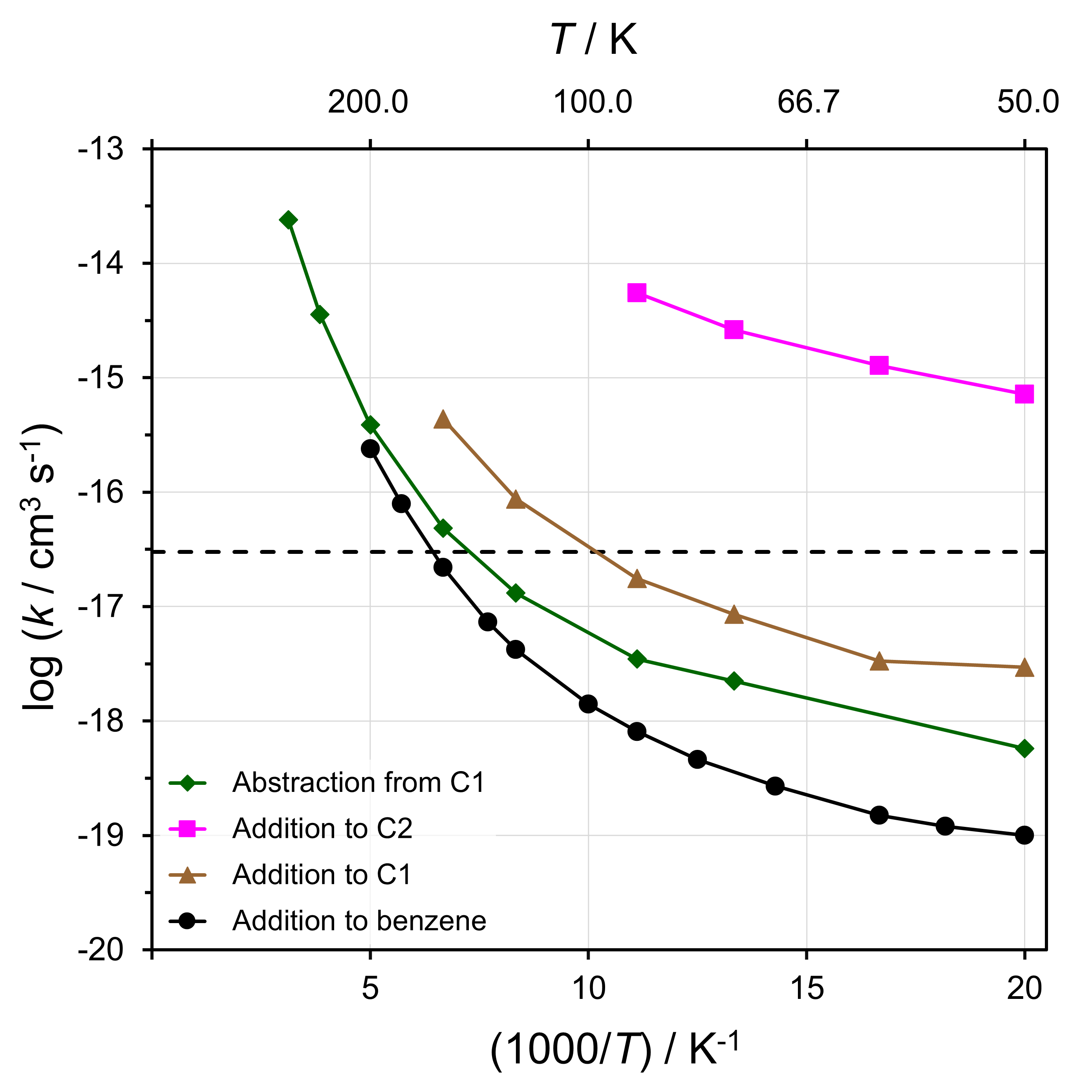}
    \caption{Temperature dependence of the \warn{rate constants} of H-atom-addition and H-atom-abstraction reactions for cyclopentadiene, as obtained at the MPWB1K/cc-pVTZ level of theory. For the H-atom-addition on benzene, MPWB1K/def2-TZVP data by Miksch et al. \cite{miksch2021} is shown for comparison. The horizontal dashed line denotes the estimated formation rate of H$_{2}$ in PDRs, as suggested by Habart et al. \cite{habart2004}.}
    \label{fig:cp_rate}
\end{figure}

\begin{figure}[H]
    \centering
    \includegraphics[width=0.5\textwidth]{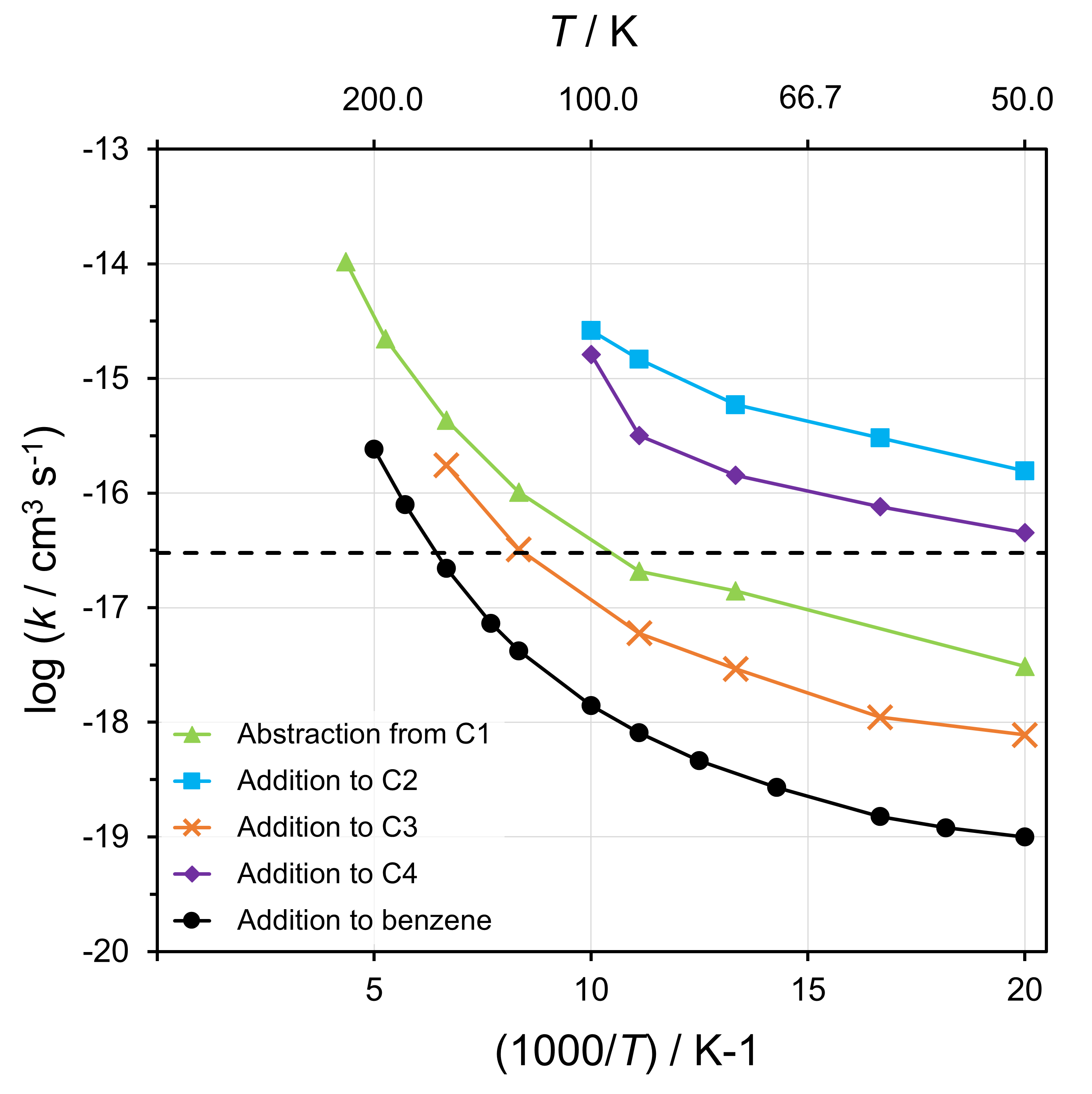}
     \caption{Temperature dependence of the \warn{rate constants} of H-atom-addition and H-atom-abstraction reactions for cycloheptatriene, as obtained at the MPWB1K/cc-pVTZ level of theory for the upper reaction paths (see text). For the H-atom-addition on benzene, MPWB1K/def2-TZVP data by Miksch et al. \cite{miksch2021} is shown for comparison. The horizontal dashed line denotes the estimated formation rate of H$_{2}$ in PDRs, as suggested by Habart et al. \cite{habart2004}.}
    \label{fig:cht_rate}
\end{figure}

Figures \ref{fig:cp_rate}  and \ref{fig:cht_rate} show the temperature dependence of the \warn{rate constants} of H-atom-addition and H-atom-abstraction reactions for cyclopentadiene and cycloheptatriene, respectively, at the MPWB1K/cc-pVTZ level of theory. 
For comparison, the respective data for the H-atom-addition reaction of benzene, as computed by Miksch and co-workers \cite{miksch2021} at the MPWB1K/def2-TZVP level of theory, is also displayed.
\warn{First of all, it should be noted that in contrast to the MPWB1K/cc-pVDZ level of theory calculations, at the MPWB1K/cc-pVTZ level of theory, all considered reactions are faster at each temperature than the H-atom-addition of benzene.
Furthermore,} as can be seen in the figures, the curves have a similar shape, but they are slightly diverging: while at around 200 K the rate of the fastest reaction is ca. two orders of magnitude larger than that of the H-atom-addition reaction of benzene, at 50 K this difference is slightly more than three orders of magnitude.
This result indicates that with decreasing temperature the defects in the aromaticity have an increasing role in H-atom-addition and H-atom-abstraction reactions, hence also in the catalytic effect in the interstellar H$_{2}$ formation reaction.

%Nincs megemlítve, hogy jönnek a sebességek a két reakcióútból össze a cikloheptatrién esetén

\subsection{Correlation between barrier heights, HOMA indices, and \warn{rate constants}}

In Table \ref{tab:50K_instanton} the change of the HOMA indices upon the reaction, the reaction barriers and the reaction enthalpies at 0 K are also listed.
It can be seen from the data that for each H-atom-addition reaction the HOMA index, and thus the aromaticity decreases.
The largest reduction is obtained for benzene.
For H-atom-abstraction reactions from the sp$^3$ carbon, the HOMA index and thus the aromaticity increases.
However, no strong correlation between $\Delta$HOMA values and the log $k$ values can be observed. 
This might be explained by the fact that besides electronic effects, steric effects also affect the \warn{rate constants} to a notable extent.
The correlation between the barrier heights and the log $k$ values is much better, with a correlation coefficient of \warn{0.88} (or \warn{0.96}, excluding H-atom-abstraction reactions), although this value is not very robust due to the limited sample size.
This not perfect, but moderate correlation is due to the fact that we investigated similar H-atom-abstraction and H-atom-addition reactions, for which similar barrier widths can be expected.
Nevertheless, from these data we can conclude that the change in the simple HOMA index is not sufficient to predict the relative \warn{rate constants}, better predictions can be obtained by the computation of the barrier heights. 
For a more reliable estimation of the reaction rate, however, the more expensive instanton model computations are indispensable.

%%-----------------------------------------------
\section{Astrochemical relevance}

The present computational results clearly indicate that, similarly to heteroatoms in heterocycles \cite{miksch2021}, the defects in the aromaticity, i.e., sp$^3$ carbon atoms, substantially increase the catalytic effects of PAHs in the interstellar H$_{2}$ formation.
In this work, we have studied small five- and seven-membered rings, such as cyclopentadiene and cycloheptatriene, and bicyclic and tricyclic systems, such as indene and 1H-phenalene.
It is shown that compared to the catalytic effect of benzene these molecules can increase the H$_{2}$ formation rate by four orders of magnitude around 50 K.
The rate computed for these reactions are larger than the low-estimate of H$_{2}$ formation rate in PDRs \cite{habart2004}.
Cyclopentadiene and indene were recently observed in the ISM \cite{cernicharo2021, burkhardt2021}, and these species might contribute to the H$_{2}$ formation themselves.
More generally, the structures studied in this paper are important structural motifs both in top-down syntheses of PAHs from graphenes, or in the bottom-up formation of larger PAHs and fullerenes from small PAHs or acetylene \cite{zhao2020}, therefore it can be expected that various PAHs with similar aromaticity defects are present in the ISM, and can significantly increase the H$_{2}$ formation rate.

%% classes.  Rather than use \section, an appropriate macro is
%% provided that will always work.
%%%%%%%%%%%%%%%%%%%%%%%%%%%%%%%%%%%%%%%%%%%%%%%%%%%%%%%%%%%%%%%%%%%%%
\section*{Acknowledgements}
This paper is dedicated to Professor Péter G. Szalay on the occasion of his 60th birthday. 
The support of the Lendület program of the Hungarian Academy of Sciences is acknowledged. This work was also supported by the ELTE Institutional Excellence Program (Grant TKP2021-NKTA-64).
%and by the National Research, Innovation and Development Fund (NKFIA) Grant No. \redx 
The authors acknowledge Dr. Yair Litman for helpful discussions about the ring-polymer instanton technique. The authors appreciate the computational resources provided by the ELTE IIG High-Performance Computing facility. 

\section*{Conflict of interest}
The authors declare no potential competing interest.

\section*{Disclosure statement} 

No potential conflict of interest was reported by the author(s). 

\section*{Data availability statement} 

\supp: \warn{Tables S1-S11: The rate constants, the number of the beads, and the crossover temperatures corresponding the instanton geometries of species discussed in the paper.
Table S12:} MPWB1K/cc-pVTZ structures (Cartesian coordinates in Å), energies and computed
harmonic vibrational frequencies (in cm$^{-1}$) and \warn{IR intensities} (km mol$^{-1}$) of species discussed in the paper. The research data supporting this publication can be accessed at https://doi.xxxx 

\section*{ORCID} 

Dávid P. Jelenfi https://orcid.org/0000-0002-6806-644X

Anita Schneiker https://orcid.org/0000-0002-8399-1529

Attila Tajti https://orcid.org/0000-0002-7974-6141

Gábor Magyarfalvi https://orcid.org/0000-0002-1464-7249

György Tarczay https://orcid.org/0000-0002-2345-1774

%\begin{suppinfo}
%???? Tables with the date of the figures????
% \end{suppinfo}


%\bibliographystyle{achemso}       % ACS ?
%\bibliographystyle{mybst}      
\bibliographystyle{tfo.bst}
%\bibliography{all,local}   % name your BibTeX data base

\end{document}